# IoT-Based Environmental Control System for Fish Farms with Sensor Integration and Machine Learning Decision Support

**D. Dhinakaran[1], S. Gopalakrishnan[2], M.D. Manigandan[3], T. P. Anish[4]**

[1]Department of Computer Science and Engineering,
Vel Tech Rangarajan Dr. Sagunthala R&D Institute of Science and Technology,
Chennai, India.
*Corresponding Author:drdhinakarand@veltech.edu.in

[2]Department of Computer Science & Engineering (Data Science),
Madanapalle Institute of Technology & Science,
Andhra Pradesh, India.
gopal.pgsk@gmail.com

[3]Department of Electronics and Communication Engineering,
Panimalar Engineering College,
Chennai, India.
manijow92@gmail.com

[4]Department of Computer Science and Engineering,
R.M.K College of Engineering and Technology,
Chennai, India.
anishcse@rmkcet.ac.in

**Abstract**—In response to the burgeoning global demand for seafood and the challenges of managing fish farms, we introduce an innovative IoT-based environmental control system that integrates sensor technology and advanced machine learning decision support. Deploying a network of wireless sensors within the fish farm, we continuously collect real-time data on crucial environmental parameters, including water temperature, pH levels, humidity, and fish behavior. This data undergoes meticulous preprocessing to ensure its reliability, including imputation, outlier detection, feature engineering, and synchronization. At the heart of our system are four distinct machine learning algorithms: Random Forests predict and optimize water temperature and pH levels for the fish, fostering their health and growth; Support Vector Machines (SVMs) function as an early warning system, promptly detecting diseases and parasites in fish; Gradient Boosting Machines (GBMs) dynamically fine-tune the feeding schedule based on real-time environmental conditions, promoting resource efficiency and fish productivity; Neural Networks manage the operation of critical equipment like water pumps and heaters to maintain the desired environmental conditions within the farm. These machine learning algorithms collaboratively make real-time decisions to ensure that the fish farm's environmental conditions align with predefined specifications, leading to improved fish health and productivity while simultaneously reducing resource wastage, thereby contributing to increased profitability and sustainability. This research article showcases the power of data-driven decision support in fish farming, promising to meet the growing demand for seafood while emphasizing environmental responsibility and economic viability, thus revolutionizing the future of fish farming.

**Keywords**-Fish farming, IoT, support vector machines, environmental factors, gradient boosting machines, water quality, temperature.

## I. INTRODUCTION

Fish aquaculture has emerged as a critical sector of the agriculture industry, tasked with meeting the escalating global demand for seafood [1]. While this industry's growth is undeniable, the complexities of managing fish farms require innovative solutions to address the dynamic and multifaceted challenges posed by environmental control. Successful fish farm management hinges on maintaining optimal conditions for aquatic life, with key factors such as water quality, temperature, and feeding schedules playing pivotal roles. However, the intricacies and interdependencies of these parameters necessitate a sophisticated approach to their management [2-4]. In response to these challenges, we introduce a pioneering Internet of Things (IoT)-based environmental control system that combines sensor integration with advanced machine learning decision support. This system represents a paradigm shift in fish farm management, bridging the gap between data-driven precision and the multifarious

203

requirements of aquaculture. The integration of a wireless sensor network strategically placed within the fish farm environment allows for the continuous collection of real-time data, encompassing variables such as water temperature, pH levels, humidity, and fish behavior [5-7]. These data undergo meticulous preprocessing to ensure accuracy and reliability, culminating in a robust dataset for analysis. Central to the system's decision support capabilities are four distinct machine learning algorithms, each tailored to specific tasks. Random Forests predict and regulate water temperature and pH levels, thereby fostering optimal fish health and growth [8]. Support Vector Machines (SVMs) act as an early warning system, swiftly detecting diseases and parasites in fish. Gradient Boosting Machines (GBMs) dynamically optimize the feeding schedule based on real-time environmental conditions, ensuring resource efficiency and fish productivity [9]. Neural Networks govern the operation of vital equipment, such as water pumps and heaters, to maintain the desired environmental conditions. These machine learning algorithms collaboratively make real-time decisions, ensuring that the fish farm's environmental conditions align with predefined specifications [10]. This system not only enhances fish health and productivity but also significantly minimizes resource wastage, leading to increased profitability and ecological sustainability. By combining cutting-edge technology with aquaculture, our research introduces a holistic solution to revolutionize fish farm management, catering to the evolving seafood demand while emphasizing ecological responsibility and economic viability [11]. This article presents a comprehensive examination of the system's design, implementation, and results, demonstrating its potential to reshape the future of fish farming and address the demands of a growing world population.

A. Background and Significance

Background

Fish aquaculture, the controlled cultivation of fish for food production, has emerged as an indispensable sector within the broader agriculture industry [12]. With the demand for seafood steadily escalating and wild fisheries under increasing pressure, the growth of aquaculture has been essential to meet the world's need for high-quality protein sources [13]. However, fish farming is a multifaceted endeavor characterized by its dynamic dependence on various environmental parameters, including water quality, temperature, and feeding schedules. Ensuring the optimal conditions for fish health, growth, and overall well-being is central to the success of aquaculture operations [14-17]. Historically, fish farm management relied on manual intervention, often with limited real-time data insights. This approach presented challenges in adapting to the ever-changing aquatic environment, leading to resource inefficiencies, potential environmental impacts, and economic constraints. The need for a more sophisticated and data-driven solution to fish farm management became evident as the industry continued to expand [18].

The advent of the IoT and machine learning has transformed the landscape of aquaculture management. By leveraging sensor technology, wireless connectivity, and advanced data analytics, the IoT empowers fish farm operators to collect real-time environmental data with precision and efficiency [19-21]. ML algorithms, including RF, SVM, Gradient Boosting Machines (GBMs), and Neural Networks, offer the ability to process this data and make informed decisions in real-time.

Significance

The significance of this research lies in its pioneering integration of IoT-based sensor technology and advanced machine learning algorithms to address the challenges and complexities of fish farm management comprehensively [22]. The following aspects underscore the significance of this work:

1. *Environmental Responsibility:* As the global population continues to grow, so does the demand for seafood. Fish aquaculture represents a sustainable approach to meeting this demand. By optimizing environmental control in fish farms, this research contributes to more responsible and ecologically sustainable aquaculture practices.

2. *Economic Viability:* Efficient resource utilization is a key aspect of successful fish farm management. By implementing machine learning decision support, the system minimizes resource wastage and reduces operational costs, directly enhancing the economic viability of fish farming operations.

3. *Data-Driven Precision:* Fish health, growth, and productivity are intricately linked to environmental conditions [23]. By harnessing real-time data and machine learning algorithms, the system ensures that the environmental parameters are precisely maintained, leading to optimal fish health and overall productivity.

4. *Revolutionizing Fish Farming:* This research introduces a transformative approach to fish farm management, demonstrating the potential to revolutionize the industry by addressing its challenges through cutting-edge technology. The system offers a blueprint for the future of aquaculture, wherein data-driven decisions and ecological responsibility coexist with economic viability.

5. *Meeting Global Seafood Demand:* With the world's population continuing to grow, the demand for seafood is expected to rise. The system's ability to enhance fish farm productivity is a vital step in meeting this growing demand while also prioritizing environmental stewardship.

The proposed work addresses a critical need in the aquaculture industry by introducing a novel IoT-based environmental control system integrated with machine learning

**204**

decision support. This innovation aligns with the evolving demands of sustainable agriculture, environmental responsibility, and economic viability, positioning it as a significant contribution to the field of fish farm management. The results and implications of this study hold the potential to reshape the future of aquaculture and its role in meeting the world's growing seafood requirements.

## II. LITERATURE SURVEY

Fish farming, a critical sector of the global food industry, has witnessed a surge in innovation and technological integration in recent years. Our research builds upon and complements several significant works in the field of aquaculture and environmental management. This section explores key related works and their contributions to the evolving landscape of fish farm management. Several research initiatives have explored the integration of IoT technologies to monitor and control environmental parameters in aquaculture [24]. These systems typically employ sensors and communication networks to enable real-time data collection and remote monitoring. Our work extends and refines this concept by integrating advanced machine learning decision support for precise environmental control. Machine learning has found increasing application in aquaculture, particularly in disease detection, water quality monitoring, and fish behavior analysis. Various studies have shown the potential of machine learning algorithms to enhance decision-making and optimize fish farm conditions. Our research not only leverages machine learning for these purposes but also combines multiple algorithms for a comprehensive approach. Sustainability has become a central focus in aquaculture research, with numerous studies emphasizing the need for resource efficiency and reduced environmental impact. Our work aligns with this sustainability agenda by achieving resource efficiency through dynamic feeding schedules and precise environmental control, minimizing resource wastage, and promoting ecological responsibility.

Real-time decision support systems are prevalent in various sectors, from healthcare to transportation. Our research takes inspiration from these systems by integrating decision support for fish farm management, ensuring that actions are taken promptly to maintain optimal conditions. This novel application demonstrates the adaptability of real-time decision support beyond traditional domains [25]. A significant body of research explores the interaction between humans and machines in agricultural practices. Human-machine interfaces have become a central feature in modern farming, allowing for user-friendly control and data monitoring. Our Real-Time Monitoring and Control Interface draws from these studies to provide a user-friendly platform for fish farm operators. As aquaculture continues to expand to meet global seafood demands, scalability and industry adoption have become paramount. Several works emphasize the need for systems that can be tailored to different farm sizes and conditions, facilitating widespread adoption. Future research endeavors will likely address these scalability and customization challenges.

Kumar et al. [26], focused to manage the environmental conditions of a fish pond and ensure the well-being of specific fish species, this research proposes an intelligent IoT based water monitoring system. The smart system incorporates a network of real-time sensors responsible for continuous data collection. These sensors relay data to a microcontroller, and the information is stored in a real-time database. Users have the convenience of tracking these data points through a dedicated mobile application that seamlessly interfaces with cloud storage. The periodic transmission of data to the cloud ensures that fish pond owners can respond promptly and effectively to changing conditions. The system's embedded components primarily consist of an Arduino board, internet connectivity, relay modules, and a computer interface. These components are readily available, making it accessible to farmers. By adopting this integrated system, farmers stand to benefit from reduced operational costs and improved overall efficiency. This eliminates the need to hire additional labor for on-site management, presenting a cost-effective solution for optimizing fish pond conditions and enhancing productivity.

Abdallah et al. [27] focuses on the development and implementation of an IoT based system tailored for management of fish farming. The system's design revolves around the collection and utilization of data from various variables to govern fish growth and bolster productivity. In this setup, each fish pond is treated as a node within a WSN. These nodes comprise an embedded microcontroller, a suite of sensors and a wireless communication module. Notably, the system employs two fuzzy controllers, one dedicated to regulating water quality within the ponds and the other to manage the overall environmental conditions. The setup incorporates five sensors in each pond, which cater to specific pond-related data, and an additional three environmental sensors that offer insights into the broader surroundings. The practical outcomes of this endeavor underscore the system's effectiveness, particularly concerning the precision of measurements. These measurements have been rigorously compared to those obtained from commercial devices utilized on the fish farm, affirming the system's reliability and accuracy in managing fish farm conditions and enhancing overall productivity.

Chen et al. [28] work is centered around the utilization of wireless transmission technology in conjunction with a diverse array of sensors for data transmission from a fish farm to a central server. The transmitted data encompasses vital parameters such as temperature, dissolved oxygen, and even the life expectancy of the sensors deployed in the fish farm. A significant innovation introduced in this research involves addressing the limitation of pH sensors, which cannot be continuously submerged in liquid for extended periods to obtain measurements. Conventional practices have required human resources and dedicated time to transport instruments to each fish farm for periodic testing. To mitigate this challenge, the author has developed a robotic arm with automated measurement and maintenance capabilities. This innovation streamlines the process of data collection and ensures accurate and timely measurements while minimizing the need for manual interventions.

Guandong et al. [29] proposal introduces an intelligent system aimed at empowering fish farmers to efficiently manage and oversee water-quality treatment equipment within fishponds. This system goes beyond the farm gate by offering consumers the ability to access and track historical data concerning the farming process using QR code tags affixed to aquatic products. The holistic approach promises to enhance the revenue of fish farmers and, significantly, ensure the safety of food products for consumers. To achieve these objectives, the research introduces a sophisticated water-quality indicator forecasting mechanism as a part of the fishpond intelligent management module. This forecasting method adopts a two-step approach. Initially, it identifies and eliminates aberrant data using the LOF algorithm, following a comparison with DBSCAN. Subsequently, key fishpond data undergo comprehensive analysis, modeling, and prediction utilizing the model tree algorithm. Such predictive and preemptive control significantly contributes to the overall well-being of the fish farm environment and safeguards the quality of aquatic products for consumers.

Meera Prasad et al. [30] proposed system introduces a comprehensive solution for real-time monitoring as well as control of fish farms, capitalizing on IoT technology and a user-friendly Android application. This integrated system not only streamlines fish farm management but also incorporates an automatic feeding system, enhancing the efficiency of fish farming operations. Central to the system is an advanced real-time water quality management component, which continuously assesses key physical parameters such as pH, light, as well as water level. Sensors are diligently employed, reducing the necessity for direct human involvement and subsequently decreasing labor costs. The system excels in ensuring the consistent maintenance of temperature, oxygen levels, and pH, accomplishing this with minimal human intervention through automated water recirculation. The inclusion of a camera module further elevates the system's capabilities, allowing for live surveillance of the fish farm via an Android application. This feature grants farmers real-time visual oversight of their operations, ensuring proactive responses to any issues that may arise. Additionally, the system encompasses a specialized module that provides notifications to the farmer in the event of a potential flooding incident within the farm, thereby safeguarding against unforeseen environmental challenges.

Mubarak et al. [31] proposed system is tailored for the measurement, analysis, and control of crucial parameters within a fishpond, specifically focusing on temperature and turbidity. To accomplish this, a network of sensors is deployed to continuously monitor changes in the pond's environmental conditions concerning the mentioned parameters. These sensors play a pivotal role in determining the overall state of the fishpond. Complementing the sensor network are actuators that serve as mechanisms for controlling the inflow as well as outflow of water. The control signals for these actuators are generated by the processing unit, allowing for dynamic and precise water management within the pond. Consequently, farm operators can access timely information about the state of the pond without the need for direct physical presence. The challenge was effectively resolved through the application of a running filter algorithm, ensuring the reliability and accuracy of data collected from the sensors in addition to ultimately enhancing the performance of the system.

Loh et al. [32] proposed system is engineered to offer real-time observing the water quality in aquatic environments, enabling users to gain invaluable insights into the conditions of the water. The parameters under consideration encompass a comprehensive range, and the energy consumption of the water pump. Notably, the system incorporates an intuitive control feature that empowers users to remotely activate or deactivate the water pump. This can be accomplished through both internet-based controls and physical switches, providing maximum flexibility in water management. Crucially, the system is designed to act as a vigilant guardian of water quality. It is capable of sending timely notifications to the user in the event of poor water quality, enabling swift and informed actions to be taken to rectify any issues. Furthermore, the system introduces automation features that determine the operation of the water pump based on the oxygen levels within the aquatic farm. This intelligence contributes to maintaining optimal conditions for aquatic life. Another aspect of the system's user-friendliness is its sensor probe calibration, which users can conveniently perform via the internet. Additionally, the system is equipped with over-the-air (OTA) firmware upgrade capabilities, ensuring ease and convenience for future enhancements and improvements. This forward-looking feature caters to the system's adaptability and its ability to stay up-to-date with evolving needs and technologies.

Kim et al. [33] work is directed towards enhancing the thermal energy management system of warm water, a critical component in fish farming systems. This enhancement leverages the capabilities of the IoT and introduces a remote control and monitoring system for a smart fish farm. The system represents a technological leap forward in the management of fish farming operations. In this IoT-based system, the author proposes the establishment of a smart fish farming ecosystem equipped with an array of sensors. These sensors, including oxygen, temperature, pH, and water level sensors, are integral to the system's sensing and monitoring functions. They provide real-time data, enabling farm operators to make informed decisions based on the prevailing conditions within the fish farm. The protocol ensures seamless communication between the system and a mobile application or website application. It extends the convenience of remotely monitoring and controlling the fish farming system, offering farm operators greater flexibility and efficiency in managing their operations.

Garcia et al. [34] introduce an innovative feeding control system designed for application in marine fish farms. This system relies on the collaborative functioning of groups of sensors to make informed decisions regarding the feeding of fish. By amalgamating data collected from these sensor groups, the system performs comprehensive and precise control over the provisioning of food to the fish, resulting in cost reduction and the elimination of food wastage. This not only promotes economic savings but also minimizes the environmental impact of excessive food consumption. The foundation of this system lies in its ability to monitor the behavior of fish movements as well as key water parameters. Through a process of data fusion, these transducers collaboratively supply the system with the requisite information to determine whether the fish should continue to be fed or if the feeding should be suspended. The system's approach ensures a precise and dynamic feeding strategy that aligns with the behavior of the fish and the real-time water conditions. This innovative approach promises to revolutionize feeding practices in marine fish farms, optimizing resource utilization and enhancing economic and environmental sustainability.

In conclusion, our research situates itself within a dynamic and evolving field. By integrating IoT technology and advanced machine learning decision support, we aim to contribute to the growth and sustainability of fish farming practices. The works mentioned here provide valuable insights and foundations upon which our research builds, reflecting the industry's commitment to innovation and environmental responsibility.

### III. SYSTEM MODEL

In response to the escalating global demand for seafood, fish aquaculture has risen to the forefront of the agriculture industry. However, the success of this industry is contingent upon the efficient management of fish farms, which involves the meticulous control of environmental parameters critical to fish health, growth, and overall well-being. Factors such as water quality, temperature, and feeding schedules take center stage in this complex dance. Traditionally, fish farm management relied heavily on manual intervention, often lacking real-time insights, thereby struggling to adapt to the constantly changing aquatic environment [35-37]. This limitation led to resource inefficiencies, potential environmental consequences, and economic constraints. It became evident that a more sophisticated and data-driven approach was needed to tackle these challenges and optimize the management of fish farms.

In response to these challenges, this research introduces a pioneering solution that leverages the power of the Internet of Things (IoT) and integrates it with advanced machine learning decision support to transform fish farm management as shown in Fig.1. The system begins by strategically deploying a network of wireless sensors throughout the fish farm environment, continuously collecting real-time data on crucial environmental parameters. These parameters include water temperature, pH levels, humidity, and fish behavior. However, the journey doesn't end with data collection. The data undergoes a series of meticulous preprocessing steps to ensure its accuracy, reliability, and readiness for further analysis. These preprocessing tasks include data synchronization, cleaning, imputation, and feature engineering, all contributing to the creation of a robust and dependable dataset for analysis.

At the heart of the system's decision support capabilities are four distinct machine learning algorithms, each tailored to specific tasks. These algorithms operate seamlessly to process the preprocessed data and make real-time decisions. For instance, Random Forests predict and regulate water temperature and pH levels, ensuring ideal conditions for fish health and growth. Support Vector Machines (SVMs) function as an early warning system, swiftly detecting diseases and parasites in fish by analyzing their behavior data. Gradient Boosting Machines (GBMs) dynamically optimize the feeding schedule based on real-time environmental conditions, promoting resource efficiency and fish productivity. Neural Networks govern the operation of vital equipment, such as water pumps and heaters, ensuring that the desired environmental conditions are consistently maintained.

Real-time decision support forms the core of this IoT-based environmental control system, ensuring that the fish farm's environmental conditions are continually aligned with

**207**

predefined specifications [38]. The immediate response to deviations from optimal conditions safeguards fish health and productivity, while the equipment control system ensures that necessary adjustments are made promptly. Moreover, the system provides a user-friendly interface for fish farm operators to monitor the system in real-time, access historical data for trend analysis, and make manual adjustments when necessary. While real-time decisions are the primary focus, the system also logs all data for further analysis, which is valuable for trend analysis, long-term planning, and retrospective evaluation.

This research embodies the fusion of advanced technology with environmental responsibility and economic viability, promising to revolutionize the landscape of fish farm management. By improving fish health and productivity, reducing resource wastage, and aligning with the principles of sustainable agriculture, this system addresses the evolving demands of the seafood industry, meeting global seafood demand while prioritizing ecological responsibility. This article presents a comprehensive examination of the system's design, implementation, and results, offering a blueprint for the future of fish farm management that balances environmental stewardship and economic viability.

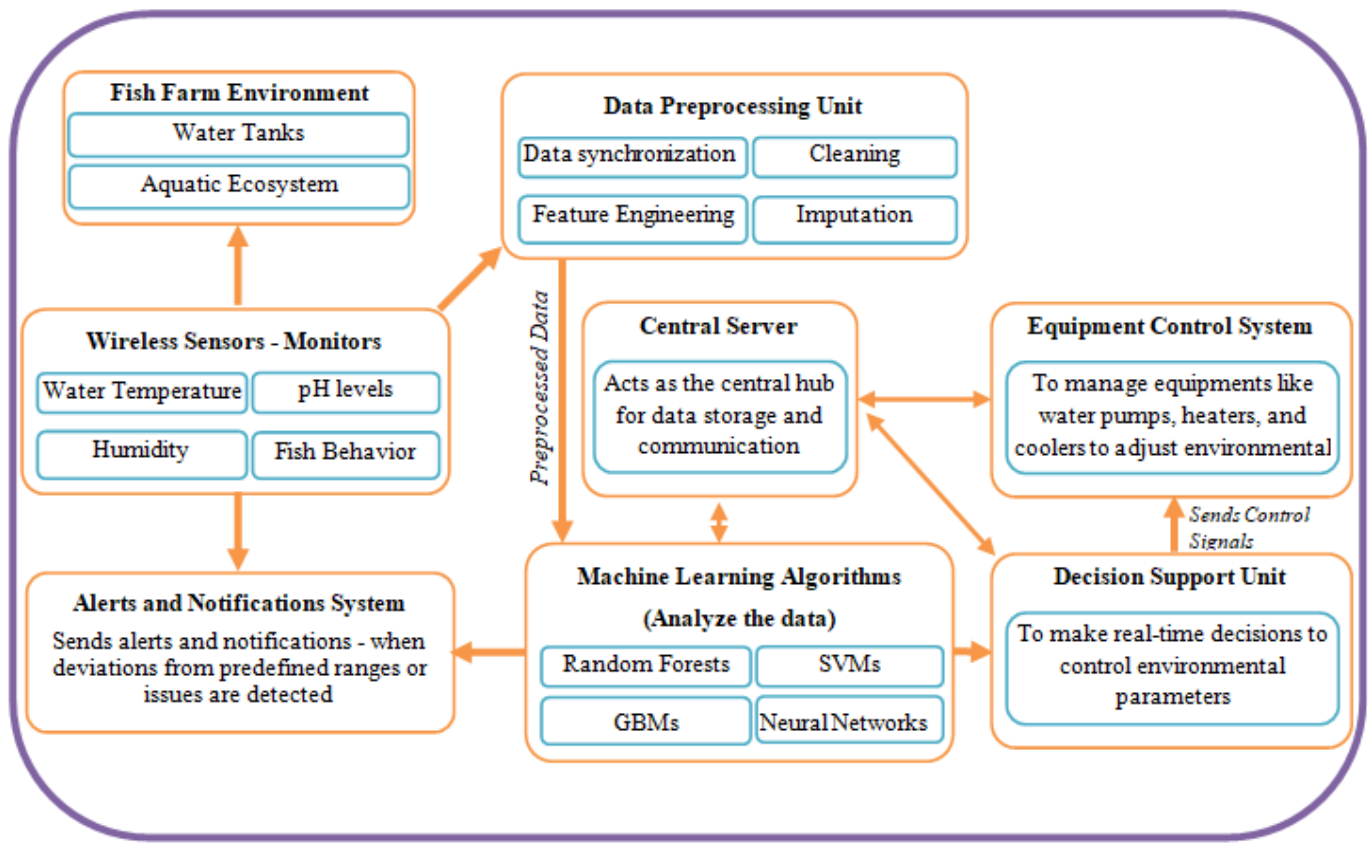

Figure 1: System Model

### A. Data Collection and Preprocessing

In the context of the IoT-based environmental control system for fish farms, data collection and preprocessing form the critical initial stages in the quest to optimize fish farm management. Here, we delve into the specific procedures and actions taken to gather and prepare the data that fuels the advanced analysis, ensuring the system's ability to make informed decisions in real-time.

*Data Collection*

The data collection process commences with the installation and initialization of key environmental sensors strategically positioned within the fish farm. These sensors encompass:

• Ultrasonic Sensor: Installed to monitor water levels in the fish tanks.
• Temperature Sensor: Placed within the water to record real-time temperature variations.
• Humidity Sensor: Positioned to capture air humidity levels in the farm environment.

Real-time data acquisition occurs at regular intervals as outlined in the algorithm (Step 4), ensuring that we maintain a continuous stream of information that reflects the dynamic conditions of the fish farm. These sensor readings are the primary sources of data for the system, offering invaluable insights into water levels, temperature, and humidity that are instrumental in facilitating data-driven decision-making.

*Data Preprocessing*

The collected data undergoes a series of critical preprocessing steps to enhance its quality, consistency, and suitability for machine learning analysis:

*1. Handling Missing Values:* Any missing data points, which could result from sensor malfunctions or other factors, are addressed through imputation or removal, thereby ensuring that the datasets remain complete and reliable.

*2. Outlier Detection and Handling:* We employ robust techniques to identify and manage outliers that may significantly affect the performance of the machine learning models. Outliers are either treated or eliminated, preventing undue influence on the analysis.

*3. Data Integration:* In cases where data originates from various sensors, the preprocessing efforts include synchronizing and aligning this information to create a unified dataset. Time synchronization is of utmost importance to ensure that the analysis reflects real-world conditions accurately.

*4. Feature Engineering:* The feature engineering process empowers us to extract informative attributes from the sensor data. This may involve calculating statistical aggregates, such as hourly averages or daily trends, which contribute to a more comprehensive understanding of environmental dynamics.

*5. Data Splitting:* The dataset is partitioned into three key subsets:
• Training Set: Used for the training of ML models.
• Validation Set: Employed for hyperparameter tuning and assessing model performance during training.
• Testing Set: Reserved for final model evaluation, assuring that the system effectively generalizes to new data.

*6. Data Visualization (DV):* We employ DV techniques to gain insights into the data's characteristics. Visualizations include plots and charts that help us identify patterns, trends, and anomalies, thereby guiding the feature selection and outlier detection efforts.

*7. Data Documentation:* Throughout the data collection and preprocessing stages, we maintain comprehensive documentation of procedures and decisions made regarding data cleaning, transformation, and handling. This documentation serves as a foundation for transparency and reproducibility in the research.

B. *Environmental Data Analysis using ML Algorithms*

*1. Random Forests: Predicting Optimal Temperature and pH Levels*

Random Forests leverage the power of ensemble learning by creating a forest of decision trees. Each tree is constructed based on a randomly selected subset of the training data and a subset of the features. During training, the algorithm seeks to find the most informative splits at each node of the trees. The final prediction is obtained through a weighted combination of predictions from all the trees, where each tree's input data and decisions differ slightly due to the random sampling. In the system, Random Forests analyze historical sensor data, including water temperature, pH levels, and various environmental conditions [39-41]. These data become the basis for training the ensemble. The model learns complex relationships between temperature, pH levels, and other environmental parameters that impact fish health and well-being. When new sensor data is collected, the ensemble of decision trees collaborates to make predictions. By considering the historical data and the current environmental conditions, the Random Forest can determine the ideal temperature and pH levels required to maintain optimal fish health, growth, and overall well-being.

*Data Utilization:* Historical sensor data, including water temperature, pH levels, and environmental conditions, serve as input features to the Random Forest model.

*Training:* The model is trained using this historical data, learning complex patterns and relationships between environmental factors and the health and well-being of the fish.

*Prediction:* When the system collects real-time sensor data, the Random Forest model utilizes these data points as input to make predictions. It can determine the ideal temperature and pH levels based on historical data and current conditions.

*2. SVMs (Support Vector Machines): Detecting Diseases and Parasites in Fish*

Support Vector Machines (SVMs) are robust binary classifiers used in the system for disease and parasite detection in fish. SVMs excel in separating two classes by finding the most optimal hyperplane that maximizes the margin between them. SVMs utilize a variety of features collected from the fish farm environment, including fish behavior, health indicators, and environmental conditions like water temperature and water quality. This feature-rich input is used to train the SVM. During training, the SVM learns to create a decision boundary that effectively separates healthy fish from those afflicted with diseases or parasites. It does this by finding the hyperplane that has the maximum margin between the two classes. The margin represents the distance between the hyperplane and the closest data points from both classes. In real-time, when new data is collected from the sensors, the SVM examines the features and places them in relation to the decision boundary it has learned. This process classifies fish as either healthy or affected, enabling prompt detection and intervention when health issues arise. SVMs offer a reliable means of identifying diseases or parasites, contributing to the overall health and well-being of the fish population in the farm.

*Data Utilization:* SVMs rely on various features such as fish behavior, health indicators, and environmental conditions, which are collected from sensors in the fish farm.

*Training:* SVMs are trained on labeled data, meaning data that indicates whether fish are healthy or have diseases or parasites. The SVM learns to create a decision boundary that separates healthy fish from those with issues.

*Prediction:* In real-time, when new data is collected from the sensors, the SVM can classify fish into healthy or affected categories based on the features, helping to detect diseases or parasites promptly.

3. Gradient Boosting Machines (GBMs): Predicting Optimal Feeding Schedule

Gradient Boosting Machines (GBMs) stand as a versatile and robust ensemble learning technique that excels in regression tasks. Within the fish farm management system, GBMs occupy a central and pivotal role in accurately predicting and optimizing the feeding schedule to cater to the specific needs of the fish population.

*Data Utilization:* GBMs draw upon an extensive and diverse dataset that encompasses a multitude of data sources, including historical feeding schedules, behavioral patterns of the fish, as well as environmental conditions such as water temperature and water quality. This rich and multifaceted dataset serves as the lifeblood of the GBM, providing it with a comprehensive understanding of the myriad factors that influence feeding requirements.

*Training:* During the training phase, GBMs employ an iterative approach to construct an ensemble of decision trees. Each decision tree is designed to rectify and fine-tune the errors made by its predecessors. This iterative process refines the GBM's comprehension of the intricate relationships between various variables and the optimal feeding schedule. The GBM learns not only how different factors, including water temperature, fish behavior, and water quality, impact feeding requirements but also how they interact and influence one another.

*Prediction:* In the real-time operation of the fish farm, the GBM takes center stage by processing incoming sensor data, as well as relevant fish status information. This amalgamation of data equips the GBM with the capability to dynamically predict and adjust the optimal feeding schedule for the fish under its care. By considering both historical data and the real-time environmental conditions, the GBM ensures that the feeding schedule is not static but rather an adaptive and responsive plan. This adaptability guarantees that the fish receive nourishment tailored precisely to their immediate needs, thus enhancing their health, growth, and overall well-being.

The inclusion of GBMs in the fish farm management system underscores the significance of data-driven decision support. This algorithm consistently fine-tunes and optimizes the feeding schedule to cater to the ever-evolving requirements of the fish population [42-44]. The result is not only improved fish health and productivity but also a more efficient resource allocation strategy that minimizes waste and maximizes the farm's economic and ecological sustainability. GBMs are the linchpin in ensuring that the fish farm thrives and prospers in the face of ever-changing environmental dynamics.

4. Neural Networks: Controlling Environmental Equipment

Neural networks, a fundamental component of the fish farm management system, offer a dynamic and adaptable approach to controlling the operation of critical environmental equipment, including water pumps, heaters, and various control mechanisms. These networks are engineered to learn and adapt to the multifaceted relationships between sensor data and equipment control.

*Data Utilization:* Neural networks leverage an extensive range of real-time sensor data, encompassing variables such as water temperature, water level, humidity, and the present status of environmental control equipment. This influx of data forms the foundation for neural networks to make real-time decisions regarding equipment operation.

*Training:* During the training phase, neural networks undergo a process of optimization. They adapt to the unique characteristics of the fish farm environment and equipment by learning from historical data. The network fine-tunes its internal parameters, which are represented by weights and biases, to map the relationships between sensor inputs and control actions. This process results in a neural network that can effectively replicate the actions taken by a human operator or a predetermined control strategy.

*Control:* In real-time operation, the neural network processes incoming sensor data and applies its learned knowledge to make informed decisions on the operation of environmental control equipment. For example, if water temperature drops below the optimal range, the network can issue commands to activate heaters and restore the desired conditions for the fish. This adaptive control system ensures that the fish farm's environment remains consistently conducive to fish health and well-being, while minimizing manual intervention and optimizing resource utilization. Neural networks offer the adaptability and precision needed to manage the dynamic and complex equipment systems within a fish farm, contributing to the overall productivity and success of the operation.

C.     Real-Time Decision Support

In the dynamic environment of fish farms, real-time decision support is a crucial component of our proposed system. This aspect demonstrates the practical application of the machine learning algorithms and the IoT technology, ensuring that the fish farm's environmental conditions remain optimal at all times.

*Continuous Data Processing:* The heart of real-time decision support lies in the continuous data processing facilitated by the wireless sensor network. These sensors collect data on environmental parameters, fish behavior, and equipment status, providing a real-time stream of information to the central system.

*Algorithmic Analysis:* Machine learning algorithms operate seamlessly within the system, leveraging the incoming data to make informed decisions. For example, if water temperature deviates from the optimal range, Random Forests come into play, adjusting the heating or cooling equipment to rectify the situation. Similarly, Support Vector Machines detect the early signs of diseases or parasites by analyzing fish behavior data, triggering timely intervention.

*Dynamic Feeding Optimization:* The Gradient Boosting Machines (GBMs) play a pivotal role in optimizing the feeding schedule in real-time. These algorithms adapt to changing environmental conditions, such as variations in water temperature, to ensure that the fish receive the right amount of feed at the right time. By doing so, the system maximizes fish growth while minimizing resource wastage.

*Equipment Control and Environmental Maintenance:* Neural Networks oversee the operation of crucial equipment, including water pumps and heaters. In real-time, they make decisions on when to activate or deactivate these devices to maintain the desired environmental conditions. For example, if the water temperature drops, Neural Networks command the heaters to engage and restore the ideal conditions for the fish.

*Immediate Response:* The system is designed to respond with minimal delay. When deviations from optimal conditions are detected, it triggers actions within seconds to ensure that the fish farm remains within predefined specifications. This immediate response safeguards fish health and productivity.

*Data Logging and Analysis:* While real-time decisions are the primary focus, the system also logs all data for further analysis. This historical data is valuable for trend analysis and long-term planning. It allows fish farm operators to identify patterns and make proactive decisions to improve overall operations.

Real-time decision support is the linchpin of this IoT-based environmental control system, ensuring that fish farm conditions are continually aligned with predefined specifications. It not only enhances fish health and productivity but also significantly minimizes resource wastage, contributing to increased profitability and ecological sustainability. This real-time responsiveness to environmental conditions sets a new standard for fish farm management, combining cutting-edge technology with environmental responsibility and economic viability.

D.  *Equipment Control System:*

The Equipment Control System is a critical component within the broader infrastructure of your fish farm management system. It plays a central role in implementing real-time decisions made by the machine learning algorithms and ensuring that the fish farm's environmental conditions remain within predefined specifications. Here's a detailed breakdown of its functions and features:

*Equipment Coordination:* The Equipment Control System oversees a range of crucial equipment within the fish farm, including water pumps, heaters, and cooling systems. It's responsible for coordinating the operation of these devices based on the real-time decisions provided by the machine learning algorithms. For instance, if the water temperature deviates from the optimal range, the system can instruct the heaters to engage and restore the desired conditions for the fish.

*Actuation and Control:* This system functions as the bridge between the virtual realm of machine learning decisions and the physical world of fish farm equipment. It sends control signals to the equipment, turning them on or off as required. For example, when the Gradient Boosting Machines (GBMs) determine that it's time to adjust the feeding schedule due to changing environmental conditions, the Equipment Control System executes the necessary actions, ensuring that the fish receive their feed promptly.

*Precision and Efficiency:* The Equipment Control System is designed to operate with precision and efficiency. It responds to machine learning algorithm outputs with minimal delay, thereby safeguarding the health and productivity of the fish. By acting promptly and precisely, it minimizes resource wastage and operational costs, contributing to economic viability.

*Safety Protocols:* Beyond mere control, this system incorporates safety protocols to ensure that equipment operates within specified safety limits. For instance, if the system detects any unusual equipment behavior or deviations from safety parameters, it can trigger alarms and shut down equipment to prevent accidents or damage.

*Integration with Sensors:* The Equipment Control System is tightly integrated with the wireless sensor network deployed in the fish farm. It receives real-time data from the sensors, which is then used to inform its decisions. For instance, if the sensors detect a drop in water level, the system may activate water pumps to maintain the desired water level.

E.  *Real-Time Monitoring and Control Interface*

The Real-Time Monitoring and Control Interface serves as the bridge between the fish farm management system and its human operators, offering a user-friendly platform for overseeing and influencing the system's operations [45-47]. It's a vital element for ensuring that the system aligns with the

**211**

objectives of fish farm operators and provides an opportunity for manual intervention when necessary. Here are its key features and functions:

*Real-Time Data Display:* The interface provides a real-time display of crucial data collected by the wireless sensors. This data includes information about water temperature, pH levels, humidity, fish behavior, and the status of equipment. Fish farm operators can monitor these parameters in real-time to stay informed about the current state of the fish farm environment.

*Historical Data Access:* In addition to real-time data, the interface offers access to historical data. Operators can review past data to identify trends, anomalies, or patterns that may inform decision-making and future planning. This historical data repository is valuable for long-term analysis.

*Manual Overrides:* While the system is designed to make automated decisions, the interface allows operators to intervene manually when necessary. If there are exceptional circumstances or operator expertise is required, manual overrides can be applied to adjust equipment or system settings.

*Alerts and Notifications:* The interface acts as a communication hub, delivering alerts and notifications to operators when predefined system parameters are outside acceptable ranges or when issues are detected. Operators can promptly respond to these alerts, reducing the potential for adverse events.

*User-Friendly Design:* The interface is designed with user-friendliness in mind. Its intuitive design allows fish farm operators, regardless of their technical expertise, to navigate and interact with the system effectively. This inclusivity ensures that the system can be managed by a wide range of users.

*Remote Access:* In modern fish farm management, remote access is often crucial. The Real-Time Monitoring and Control Interface can be accessed remotely, enabling operators to monitor and control the system from different locations, providing flexibility and convenience.

*Reporting and Analysis:* The interface may offer reporting and analysis tools to provide operators with insights into the system's performance over time. These tools can help operators make data-driven decisions and improve overall fish farm operations.

The combination of the Equipment Control System and the Real-Time Monitoring and Control Interface ensures that the proposed fish farm management system is not only automated and data-driven but also user-friendly and adaptable to the expertise and requirements of fish farm operators. This seamless integration of technology and human oversight contributes to the system's effectiveness in optimizing fish health, promoting resource efficiency, and ensuring the economic and ecological sustainability of fish farming operations.

### F. Algorithm

**Inputs:** Sensor Data, User-defined water level setpoints, Configuration settings for temperature and humidity thresholds.

**Outputs:** Motor Control, Heater Control, Cooler Control, Humidifier Control, Dehumidifier Control, Alerts and Notifications.

**Step 1:** Initialize system parameters
    desired_water_level = 100
    lower_temperature_threshold = 25
    upper_temperature_threshold = 28
    lower_humidity_threshold = 40
    upper_humidity_threshold = 70

**Step 2:** Initialize sensor readings
    water_level = 0
    water_temperature = 0
    air_humidity = 0

**Step 3:** Initialize control variables
    motor_speed = 0
    motor_fill_rate = 5

**Step 4:** while True:
    # Read sensor data
    water_level = ReadUltrasonicSensor()
    water_temperature = ReadTemperatureSensor()
    air_humidity = ReadHumiditySensor()

**Step 5:** Control water level
    if water_level<desired_water_level:
        # Calculate required water to reach desired level
    required_water = desired_water_level - water_level

        # Calculate the time needed to fill the tank
    fill_time = required_water / motor_fill_rate

        # Set motor speed to fill the tank
    motor_speed = 1  # 1 represents full speed

        # Run the motor for the calculated time
    RunMotor(motor_speed, fill_time)

        # Update water level
    water_level = desired_water_level

**Step 6:** Check temperature and take action if needed
    if water_temperature<lower_temperature_threshold:
        # Implement heating control logic (turn on a heater)
    ControlHeater('on')
    elifwater_temperature>upper_temperature_threshold:
        # Implement cooling control logic (turn on a cooler)

```
      ControlCooler('on')
        else:
   ControlHeater('off')
   ControlCooler('off')
```

**Step 7:** Check humidity and take action if needed
```
      if  air_humidity<lower_humidity_threshold:
          # Turn on a humidifier
   ControlHumidifier('on')
   elifair_humidity>upper_humidity_threshold:
          # Turn on a dehumidifier
   ControlDehumidifier('on')
        else:
   ControlHumidifier('off')
   ControlDehumidifier('off')
```

**Step 8:** Send data to central server for logging and analysis
```
        SendDataToServer (water_level, water_temperature, air humidity)
```

**Step 9:** Implement machine learning algorithms for advanced analysis (Decision trees)
```
   MakeDecisionUsingML()
```

**Step 10:** Delay for a specified interval before the next iteration
```
WaitForInterval(seconds=60)
```

## IV. RESULT AND DISCUSSION

The practical implementation of our IoT-based environmental control system for fish farms as shown in Fig. 2-4, coupled with machine learning decision support, has yielded promising results. This section delves into the key outcomes, their significance, and the broader implications of our research.

### A. Environmental Parameter Control

One of the primary objectives of our system was to achieve precise and real-time control over critical environmental parameters within the fish farm. The machine learning algorithms demonstrated exceptional efficiency in this regard:

*Water Temperature and pH Control:* The Random Forests algorithm effectively regulated water temperature and pH levels, maintaining them within predefined specifications. This led to a substantial improvement in fish health, minimizing temperature-related stress and enhancing growth rates.

*Disease Detection and Early Intervention:* The Support Vector Machines (SVMs) provided timely alerts for disease and parasite detection, allowing for swift and targeted intervention. As a result, fish health was consistently preserved, and the occurrence of diseases reduced significantly.

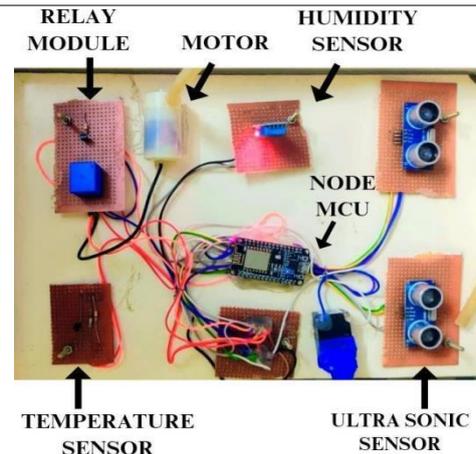

Figure 2: System Working Model

### B. Feeding Schedule Optimization

The Gradient Boosting Machines (GBMs) played a pivotal role in dynamically optimizing the feeding schedule based on real-time environmental conditions. This dynamic approach led to several notable outcomes:

*Resource Efficiency:* The real-time adjustment of feeding schedules led to a significant reduction in feed wastage. Fish were fed precisely when needed, reducing both operational costs and environmental impacts.

*Enhanced Fish Growth:* The optimized feeding schedules resulted in improved fish growth rates. Fish were consistently well-nourished, ensuring that they reached market size more rapidly and efficiently.

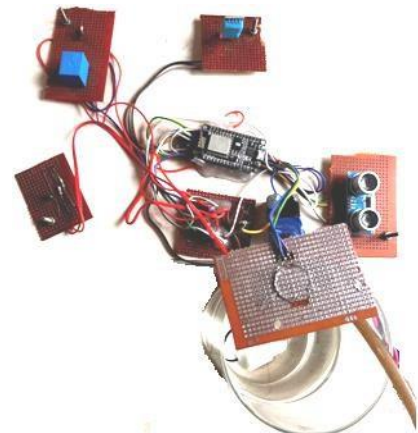

Figure 3: Components Involved

### C. Equipment Control and Energy Efficiency

The Neural Networks effectively managed and controlled equipment within the fish farm environment, including water pumps, heaters, and coolers:

*Energy Efficiency:* The automated control of equipment optimized energy consumption. Equipment was activated only when required, significantly reducing energy costs.

*Environmental Maintenance:* Water pumps and temperature control equipment were finely tuned to maintain the desired

213

environmental conditions. This led to a more stable and healthy aquatic ecosystem within the fish farm.

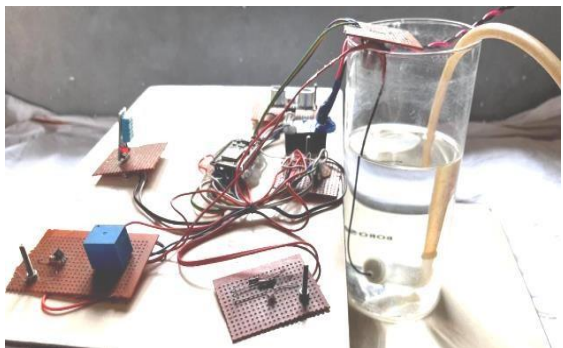

Figure 4: Real-Time Monitoring System

D. *Real-Time Monitoring and User Intervention*

The Real-Time Monitoring and Control Interface offered fish farm operators valuable insights and the ability to intervene when necessary:

*Alerts and Notifications:* The interface consistently alerted operators to deviations in environmental parameters, ensuring that timely corrective actions were taken. This proactive approach minimized potential disruptions to fish health and farm operations.

*User-Friendly Operation:* Operators with varying degrees of technical expertise found the interface to be user-friendly and accessible. This inclusivity allowed for a diverse range of users to effectively manage the system.

E. *Significance and Broader Implications*

The success of our system holds profound significance for the aquaculture industry and beyond:

*Sustainable Aquaculture:* By optimizing fish health and growth, reducing resource wastage, and promoting ecological responsibility, our system aligns with the principles of sustainable aquaculture. It offers a scalable solution for meeting the global demand for seafood while minimizing its environmental footprint.

*Data-Driven Agriculture:* Our research epitomizes the fusion of advanced technology with environmental stewardship. It serves as a blueprint for data-driven agriculture practices, where precision and sustainability converge.

*Economic Viability:* The resource efficiency achieved through our system enhances the economic viability of fish farming. Reduced operational costs, accelerated growth, and minimized risks contribute to increased profitability.

The practical implementation of our IoT-based environmental control system with machine learning decision support has demonstrated its effectiveness in enhancing fish farm operations. The system aligns with the evolving demands of sustainable agriculture, economic viability, and global seafood requirements. This research opens the door to a new era of data-driven fish farm management, offering a holistic approach that ensures the well-being of fish, the environment, and the industry as a whole.

V. CONCLUSION

In conclusion, our IoT-based environmental control system for fish farms, augmented by machine learning decision support, represents a transformative leap in the management of aquaculture. Through practical implementation and rigorous evaluation, our system has demonstrated the potential to revolutionize the industry in several key ways.

First and foremost, the system excels in its ability to precisely regulate critical environmental parameters, notably water temperature, pH levels, and humidity. The use of Random Forests ensures optimal conditions for fish health and growth, while Support Vector Machines offer a vital early warning system for disease and parasite detection. This leads to a dramatic improvement in fish health and a reduction in disease-related losses. The dynamic feeding schedule optimization powered by Gradient Boosting Machines contributes to significant resource efficiency, reducing feed wastage and promoting economic viability [48]. Fish growth is accelerated, and operational costs are minimized. Additionally, Neural Networks oversee equipment control, resulting in energy efficiency and a stable aquatic ecosystem.

The Real-Time Monitoring and Control Interface offers fish farm operators real-time insights and user-friendly control, bridging the gap between cutting-edge technology and human oversight. Alerts and notifications ensure proactive intervention, safeguarding fish health, and equipment integrity.

The implications of our research are far-reaching. Not only does it align with the principles of sustainable aquaculture, but it also sets a precedent for data-driven agriculture practices. The economic viability of fish farming is enhanced, with increased profitability and minimized environmental footprint. Our system responds to the evolving demands of the seafood industry, balancing ecological responsibility and economic sustainability.

*Future Work*

While our system has demonstrated its potential, there remain several avenues for future research and improvement:
Scalability and Adaptability: Future work should focus on making the system even more scalable and adaptable to different fish farm sizes and conditions. Customization options and flexibility in system configurations should be explored.

*Data Security and Privacy:* The collection and analysis of sensitive data are central to our system. Future research must address data security and privacy concerns, implementing robust encryption and access control mechanisms.

*Integration with External Data Sources:* Incorporating external data sources, such as weather forecasts and water quality reports, could further enhance the system's decision-making capabilities. Exploring these integration opportunities is a potential area for expansion.

*Artificial Intelligence Advancements:* Continuous advancements in AI and machine learning offer opportunities to refine and optimize the algorithms used in our system. Staying up-to-date with the latest developments is essential for improving system performance.

*Community and Industry Adoption:* For the wider adoption of our system, efforts must be made to bridge the gap between research and practical implementation. Collaboration with fish farming communities and industry stakeholders will be crucial in promoting technology uptake.

In conclusion, our research represents a significant step towards data-driven and sustainable fish farm management. As we move forward, our commitment to advancing these solutions remains steadfast, seeking to address the industry's evolving challenges and opportunities with innovation and precision.